\documentclass[twocolumn,showpacs,preprintnumbers,amsmath,amssymb,floatfix]{revtex4}

\usepackage{epsfig,psfrag}
\usepackage{dcolumn}
\usepackage{bm}

\begin{document}


\title{Superconductivity in carbon nanotube ropes: Ginzburg-Landau approach
and the role of quantum phase slips}

\author{A.~De~Martino and R.~Egger}

\affiliation{Institut f\"ur Theoretische Physik, 
Heinrich-Heine-Universit\"at,
 D-40225 D\"usseldorf, Germany}

\date{\today}

\begin{abstract}
We derive and analyze the low-energy theory of superconductivity in
carbon nanotube ropes.  A rope is modelled as an
array of ballistic metallic nanotubes, taking into account 
phonon-mediated plus Coulomb interactions, and 
Josephson coupling between adjacent tubes. We construct the
Ginzburg-Landau action including quantum fluctuations. 
Quantum phase slips are shown to cause a depression of the critical 
temperature $T_c$ below the mean-field value, and a 
temperature-dependent resistance below $T_c$.  
\end{abstract}


\maketitle

Over the past decade, the unique mechanical, electrical,
and optical properties of carbon nanotubes, 
including the potential for useful technological applications,
have created a lot of excitement.
While many of these properties are well understood
by now, the experimental observation of intrinsic 
\cite{kociak,kasnew,tang} and anomalously strong proximity-induced 
\cite{kasumov,morpurgo} superconductivity continues to pose open
 questions to theoretical understanding.   
In this paper we analyze 1D superconductivity found in ropes of
carbon nanotubes \cite{kociak,kasnew},
starting from a microscopic model of the rope in terms of an array of
individual clean single-wall nanotubes (SWNTs), 
with attractive phonon-mediated on-tube interactions 
and inter-tube nearest-neighbor Josephson couplings. 
Then the Luttinger interaction parameter $g_c$ 
and the Josephson coupling $\lambda$
describing transfer of Cooper pairs 
are crucial microscopic parameters, which, fortunately,
can be estimated rather accurately
\cite{gonzalez1,gonzalez2,ademarti}.
The coupled-chain problem corresponding
to superconducting nanotube ropes, 
where typically less than hundred metallic SWNTs
are present \cite{kociak,kasnew}, does neither permit classical
Ginzburg-Landau (GL) theory nor a standard self-consistent BCS approach,
in contrast to the situation encountered in,
e.g., quasi-1D organic superconductors 
\cite{schulz}. At this time, nanotube ropes represent wires
with the smallest number of transverse channels 
showing intrinsic superconductivity, even when
compared to the amorphous MoGe wires of diameter
$\approx 10$~nm studied in Ref.~\cite{lau}, 
where still several thousand channels are available.
Based on a microscopic derivation of the quantum GL action, we 
show that quantum phase slips (QPS's) 
\cite{tinkham,zaikin1,zaikin2,blatter} are
crucial for an understanding of experimental results
\cite{kociak,kasnew}. First, they cause a depression of the
transition temperature $T_c$ below the mean-field critical 
temperature  $T_c^0$.  
Furthermore, for $T<T_c$, a finite resistance $R(T)$ due to QPS's appears, 
which exhibits approximate power-law scaling.
Below we determine the full temperature dependence 
of $R(T<T_c)$ for arbitrary rope length.

We consider a rope  consisting of $N$ metallic
SWNTs, where disorder is assumed
to be negligible \footnote{Since about 2/3 of all SWNTs
are semiconducting, the rope contains
 $\approx 3N$ SWNTs. In our estimates, to 
take this into account, below we consider a 
reduced Josephson coupling.}.
 The validity of modelling the
rope as an array of ballistic 1D quantum wires has recently been 
discussed in Ref.~\cite{kasnew}.  
Since the K point degeneracy is inessential here, 
an individual SWNT can be described as a spin-1/2 Luttinger liquid,  
where the combined effects of Coulomb and phonon-mediated
interactions lead to an interaction parameter $g_c$ 
\cite{ademarti}, where $g_c=1$ refers to the noninteracting case
and $g_c>1$ ($g_c<1$) signals effectively attractive (repulsive)
 interactions.   
We mention in passing that lattice commensurabilities and electron-electron 
backscattering can be neglected in the intrinsically
doped SWNTs encountered in practice \cite{egger97,kane97}.
In a thick rope, Coulomb interactions are 
expected to be largely screened off, and
$g_c>1$ due to breathing-phonon exchange
\cite{ademarti}.  The only weak screening in the thinnest
ropes studied in Refs.~\cite{kociak,kasnew}
is probably linked to the absence of superconductivity in these
samples.  To probe superconductivity in ultrathin ropes, 
it is necessary to externally screen Coulomb interactions.
In principle, three different inter-tube
coupling mechanisms should now be taken into
account, namely (i) direct Coulomb interactions,
(ii) Josephson coupling (Cooper pair hopping),
and (iii) single-electron hopping.  The last process is strongly
suppressed due to the generally different chirality of 
adjacent tubes \cite{kane}, and, in addition, for $g_c>1$,
inter-SWNT Coulomb interactions can be neglected \cite{schulz}.
Therefore the most relevant mechanism is Josephson
coupling between adjacent SWNTs.

In the (idealized) rope crystal, (metallic) SWNTs of radius $R$ 
 are arranged on a trigonal lattice with 
lattice constant $a=2R+b$, where $b=0.34$~nm
\cite{kane}.
Choosing the $x$-axis parallel to the rope, and
numbering the SWNTs by $j=1,\ldots,N$, with center at
$\vec r_j=(y_j,z_j)= n_1 \vec a_1+n_2 \vec a_2$, where
$\vec a_1=a(1,0)$ and $\vec a_2=a (1/2,\sqrt{3}/2)$ span
the trigonal lattice,  allowed indices $(n_1,n_2)$ corresponding
to $j$  follow
from the condition $|\vec r_j|\leq R_{\rm rope}$.  A given 
rope radius $R_{\rm rope}$ then fixes the number of tubes $N$. The Josephson 
coupling matrix $\Lambda_{ij}$
is nonzero only for nearest-neighbor pairs $(i,j)$, where
$\Lambda_{ij}=\lambda$.  
Only singlet pairing of electrons on the same
tube is important \cite{ademarti}, leading to the on-tube order parameter 
${\cal O}_j(x,\tau)=\sum_{r\sigma}\sigma\psi_{r\sigma j}\psi_{-r,-\sigma j}$,
where $\psi_{r\sigma j}(x,\tau)$ is the electron operator for a right-
or left-moving electron ($r=\pm$) with spin $\sigma=\pm$ on the $j$th
SWNT, and $0\leq \tau<1/T$ is
imaginary time (we put $\hbar=k_B=1$). In bosonized language \cite{egger97},
\begin{equation}\label{orderpar}
{\cal O}_j= (\pi a_0)^{-1} \exp\left[i\sqrt{2\pi} 
\theta_{c,j}\right] \cos\left[ \sqrt{2\pi} \varphi_{s,j}\right],
\end{equation}
where $a_0=0.246$~nm is the SWNT lattice spacing,
$\varphi_{c/s,j}(x,\tau)$ denotes the charge/spin boson field on the $j$th
SWNT, and $\theta_{c/s,j}$ is the dual field to $\varphi_{c/s,j}$.
The Euclidean action is
\begin{equation}\label{ea}
S = \sum_{j=1}^N S_{\rm LL}[\theta_{c,j},\varphi_{s,j}] - \sum_{jk} 
\Lambda_{jk}\int dx d\tau   {\cal O}^\dagger_j {\cal O}_k^{},
\end{equation}
where the on-tube fluctuations are governed by an effective Luttinger liquid
action \cite{egger97,kane97,ademarti},
\begin{eqnarray*}
S_{\rm LL}[\theta_c,\varphi_s] & = & \int dx d\tau \Bigl\{ \frac{v_c g_c}{2} 
\left[ (\partial_\tau \theta_{c}/v_c)^2  +  
(\partial_x \theta_{c})^2 \right] \\ &+&
 \frac{v_s}{2 g_s} 
\left[ (\partial_\tau \varphi_{s}/v_s)^2  + 
(\partial_x \varphi_{s})^2 \right]  \Bigr\},
\end{eqnarray*}
with  $v_{c/s}=v_F/g_{c/s}$ for Fermi velocity $v_F=8\times 10^5$~m/sec,
 and $g_s=1$ due to spin $SU(2)$ invariance.
Note that the model (\ref{ea}) and our results below
apply beyond the specific system 
under study here, see also Refs.~\cite{schulz,carr}. 

Next we employ a Hubbard-Stratonovich transformation in order to decouple the 
Josephson term in Eq.~(\ref{ea}), 
using the complex-valued order parameter field
 $\Delta_j(x,\tau)$.  This allows
to write the partition function as
\begin{equation}\label{hs}
Z = \int {\cal D}  \Delta
\exp\left( - \sum_j S_0[\Delta_j] - \int  dx d\tau \sum_{jk} 
 \Delta^\ast_j  D_{jk} \Delta_k \right) ,
\end{equation}
where $\Delta^\ast$ is the complex conjugate field. The 
$N\times N$ matrix $D$  denotes
the positive definite part of $\Lambda^{-1}$,
since order parameter modes 
corresponding to negative eigenvalues of $\Lambda$
can never become critical.
Integration over the on-tube boson field fluctuations then leads to
\begin{equation} \label{f00}
S_0[\Delta] = -\ln \int
{\cal D}\theta_c {\cal D}\varphi_s
 e^{ -S_{\rm LL}[\theta_c,\varphi_s]- \int dx d\tau
(\Delta^\ast {\cal O} + \Delta {\cal O}^\dagger)} . 
\end{equation}
The expectation value of Eq.~(\ref{orderpar}) can 
be computed as $\langle {\cal O}_j \rangle = 
\sum_k D_{jk} \langle \Delta_k \rangle$,
where $\langle \Delta_k \rangle$ is
averaged using the action corresponding to Eq.~(\ref{hs}).

Assuming a spin gap, taking $T=0$, and allowing only
 static homogeneous configurations $\Delta_j(x,\tau)=\Delta_0$, 
Eq.~(\ref{f00}) can be evaluated explicitly
\cite{zamo}.  Saddle-point analysis then yields a relation
between the mean-field critical temperature and the $T=0$ superconducting 
gap, see Ref.~\cite{carr}.
For general order parameter $\Delta_j(x,\tau)$ or
arbitrary temperature, however, integration over the 
Luttinger phase fields in Eq.~(\ref{f00}) is impossible.
To make progress, it is instructive 
to construct the GL action \cite{nagaosa1,tinkham}, where it is crucial to
 include quantum fluctuations.  A systematic approach proceeds
via cumulant expansion of Eq.~(\ref{f00}) up to quartic order
in the expansion parameter $|\Delta|/2\pi T$ \cite{nagaosa1}.
We stress that this expansion is carried out for the single-chain 
problem, and is {\sl not}\ restricted to $N\gg 1$.
Assuming slowly varying configurations $\Delta_j(x,\tau)$,
gradient expansion yields 
the Lagrangian in Eqs.~(\ref{hs}) and (\ref{f00}) as
\begin{eqnarray}
\label{hs2} 
L & = &  \sum_{jk}V_{jk}\Delta^\ast_j \Delta_k
+ \sum_j \Bigl ( [W_1^{-1}- A] |\Delta_j|^2 \\ &+& \nonumber
 B |\Delta_j|^4 + C \left[ |\partial_x \Delta_j|^2 + 
 c_s^{-2} |\partial_\tau \Delta_j|^2 \right] \Bigr),
\end{eqnarray}
 with the Mooij-Sch{\"o}n plasma velocity \cite{mooji},
 $c_s=v_c \sqrt{\tilde{C}/\tilde{D}}$, and
$V_{jk} = \sum_\alpha (W_\alpha^{-1}-W^{-1}_1) \langle j|\alpha\rangle
\langle \alpha|k\rangle$,
where $W_\alpha$ denote the eigenvalues of $\Lambda$ in descending
order ($\alpha=1,\cdots,N$), with eigenvectors $|\alpha\rangle$,
and the $\alpha$-summation extends over 
$W_\alpha>0$ only. Furthermore, the positive coefficients $A, B, C$ are 
with $\gamma=v_c/v_s$ given by
\begin{eqnarray}\nonumber
A &=& \frac{\gamma^{g_s}}{2\pi^2 v_c}  ( \pi a_0 T/v_c )^{g_c^{-1}+g_s-2}
\tilde{A},\\
 \label{coeffs}
B &=& \frac{a_0^2 \gamma^{2g_s}}{32\pi^4 v^3_c} 
 ( \pi a_0 T/v_c )^{2g_c^{-1}+2g_s
-6} \tilde{B}, \\ \nonumber
C &=& \frac{a_0^2\gamma^{g_s}}{4\pi^2 v_c} 
 ( \pi a_0 T/v_c )^{g_c^{-1}+g_s-4} \tilde{C}.
\end{eqnarray}
Putting $g_s=1$,
dimensionless $g_c$-dependent numbers
are defined as
\begin{equation}\label{tildec}
\tilde{C}= \int dz \frac{w^2}{ f_c(z) f_s(z)}, 
\end{equation}
where  we use the notations $z=(w,u)$,
\[
 f_c(z)=|\sinh(w+iu)|^{1/g_c}, \quad 
f_s(z)=|\sinh(\gamma w+iu)|, 
\]
and $\int dz\equiv \int_{0}^{\pi} du \int_{-\infty}^\infty dw.$
 Here $\tilde{A}$ ($\tilde{D}$)
 is given by Eq.~(\ref{tildec}) with $w^2\to 1$ ($w^2\to u^2$),
and
\begin{eqnarray*}
& &\tilde B = \int \frac{dz_1 dz_2 dz_3}
{f_c(z_2) f_c(z_{13})} \Biggl[   
\frac{4}{f_s(z_2)f_s(z_{13})}-
 \frac{f_c(z_1) f_c(z_{23})}
    {f_c(z_3)f_c(z_{12})} \\ &\times&  \left( 
   \frac{f_s( z_1)f_s(z_{23})} 
{f_s(z_2)f_s( z_{13}) f_s(z_3)f_s(z_{12})}
+ (1\leftrightarrow 2) + (1 \leftrightarrow 3) \right)  
\Biggr ] ,
\end{eqnarray*}
with $z_{ij}=(w_i-w_j, u_i-u_j)$.  The quantity $\tilde{B}$
is evaluated using the Monte Carlo method, see 
also Ref.~\cite{carr}.  For $g_c=1$, we first numerically reproduced
the exact result $\tilde{B}=8\pi^2 \tilde{C}$ 
with $\tilde{C}=7\pi \zeta(3)/4$ \cite{nagaosa1}.
Numerical values can then be obtained for arbitrary $g_c$.
>From Eq.~(\ref{hs2}), previous GL results for the infinite 2D array 
of coupled 1D chains are recovered \cite{schulz,carr}.
In that case, the $V_{ij}$ term in Eq.~(\ref{hs2}) leads to transverse
gradients because $\alpha$ corresponds to transverse 
momentum $\vec k_\perp$,  with $W_\alpha^{-1}-W_1^{-1}\propto \vec k_\perp^2$.  
Note that Eq.~(\ref{hs2}) additionally includes
 quantum fluctuations and allows to describe the case
of arbitrary $N$.

\begin{figure}
\centerline{\epsfxsize=7cm\epsfysize=6cm \epsffile{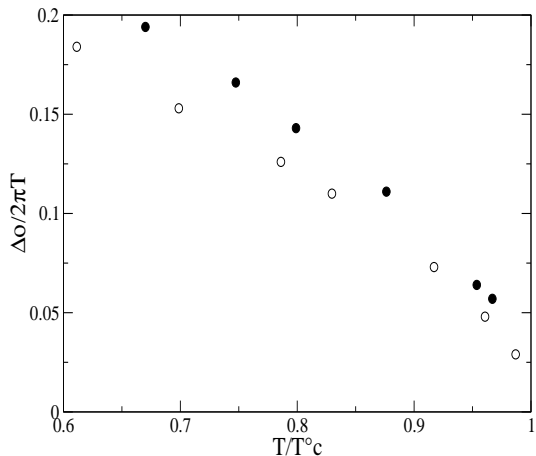}}
\caption{\label{fig1}  
Temperature dependence of $\Delta_0/2\pi T$ versus $T/T_c^0$ for
$N=31$ (open) and $N=253$ (filled circles).
}
\end{figure}

>From Eq.~(\ref{hs2}), we obtain the mean-field critical temperature
\begin{equation}\label{tc}
T_c^0 = \frac{v_c}{\pi a_0} \left( \frac{\tilde A W_1}{2\pi^2 v_F}
\right)^{g_c/(g_c-1)} .
\end{equation}
Assuming sufficiently thick ropes such that
Coulomb interactions can be neglected, in concrete estimates we shall
put $g_c=1.3$ \cite{ademarti},
with Josephson coupling $\lambda/v_F=0.02$ \cite{gonzalez2}.
Numerical evaluation yields
\begin{equation}\label{tildes}
\tilde{A} \simeq 30.72 , \quad \tilde{B} \simeq 293.1 , 
\quad \tilde{C}\simeq 12.12 ,\quad \tilde{D}\simeq 7.78 .
\end{equation}
Equation (\ref{tc}) then predicts, e.g.,  $T^0_c=2.3$~K for $N=31$,
which is slightly above reported experimental 
values \cite{kociak,kasnew}.
In what follows, we focus on temperatures below $T^0_c$.  
Writing $\Delta_j=|\Delta_j| \exp\{i\phi_j(x,\tau)\}$,
the amplitudes $|\Delta_j|$ are finite, with a gap for fluctuations
around the mean-field value.  At not too low temperatures,
they are found from the saddle point equation
\begin{equation}\label{gap}
\sum_{j} V_{ij} |\Delta_j| + (W_1^{-1}-A)|\Delta_i| +
 2B |\Delta_i|^3 =0 ,
\end{equation}
whose numerical solution (via a Newton-Raphson root finding scheme)
yields the transverse order parameter profile and 
$\Delta_0 = \sum_j |\Delta_j| /N$.
Typical results for $\Delta_0/2\pi T$ are shown in
Fig.~\ref{fig1}, which demonstrates that GL theory quantitatively holds down
to $T\approx T^0_c/2$. In our discussion below, it is useful
 even down to $T=0$.
Fixing the amplitudes $|\Delta_j|$ 
at their mean-field values, the resulting
Lagrangian governing the massless phase fluctuations (Goldstone modes) is
\begin{eqnarray}\label{final}
L&=& \sum_{j} \frac{\mu_j}{2\pi} \left[ c_s^{} (\partial_x\phi_j)^2 + c_s^{-1}
(\partial_\tau \phi_j)^2 \right]\\ \nonumber
& + & \sum_{i>j} 2V_{ij}|\Delta_i||\Delta_j| 
\cos(\phi_i-\phi_j) ,
\end{eqnarray}
with dimensionless quantities 
$\mu_j= 2\pi C |\Delta_j|^2 /c_s$.
Electromagnetic potentials can then be coupled in by standard
Peierls substitution rules \cite{nagaosa1}, e.g.~allowing
to describe the Meissner effect.
Furthermore, dissipative effects can be included following Ref.~\cite{blatter}.

In the 1D situation encountered here, superconductivity
can be destroyed by thermally activated or 
quantum phase slips \cite{tinkham}.  Following arguments similar to the ones 
of Ref.~\cite{zaikin1}, we find that only QPS's play a
role.  Numerical evaluation of Eq.~(\ref{gap}) shows
that well below $T_c^0$,
transverse fluctuations are heavily suppressed \cite{future}, and therefore 
QPS's can be described using the action
\begin{equation}\label{finala}
S=\frac{\mu}{2\pi}\int dx d\tau \left[
c_s^{-1}(\partial_\tau\phi)^2 + c_s (\partial_x \phi)^2 \right],
\end{equation} 
where $\mu=\sum_j \mu_j$ is a dimensionless rigidity.
For not too low temperature, and neglecting transverse fluctuations, 
\begin{equation} \label{mu1}
\mu(T) = \alpha_0 N \left[ 1 - (T/T_c^0)^{(g_c-1)/g_c}\right] , 
\end{equation}
where 
$\alpha_0 \simeq 4\pi \tilde{A} (\tilde{C}\tilde{D})^{1/2}  / \tilde{B}$,
resulting in $\alpha_0\simeq 12.7$ for $g_c=1.3$.  Remarkably, at
$T=0$,  Eq.~(\ref{mu1}) coincides up to a prefactor of order one
with the rigidity $\bar{\mu}$ obtained from standard 
mean-field relations \cite{nagaosa1},
$\bar{\mu}= \pi^2 n_s R^2/2 m^\ast c_s = \bar{\alpha}_0 N$, 
where $n_s$ is the density of condensed electrons. At $T=0$, this 
implies $\bar{\alpha}_0\approx v_F/c_s$.  We conclude that
the GL prediction (\ref{mu1}) for $\mu(T)$ is robust and useful even outside 
its validity regime.

QPS's are topological vortex-like excitations of the
superconducting phase field $\phi(x,\tau)$.
For rope length $L\to \infty$ and thermal length 
$L_T=  c_s/ \pi T \to \infty$, 
a QPS with core at $(x_i,\tau_i)$ and winding number $\nu_i=\pm 1$ 
(higher winding numbers are irrelevant) is
$\phi(x,\tau)=\nu_i \tan^{-1} [(x-x_i)/c_s(\tau-\tau_i)]$ \cite{chaikin},
where the finite $L,L_T$ solution follows by 
conformal transformation.  This form solves the
equation of motion for Eq.~(\ref{finala}) with a singularity
at the core, where superconducting order is locally destroyed.  
The local loss of condensation energy density $E_c$ 
(this may also contain other energy costs \cite{zaikin1}) 
leads to the core action $S_c=\kappa^2 E_c/c_s$, with 
core radius $\kappa$ as variational parameter. 
The optimal value of the core radius is $\kappa=(c_s\mu/2E_c)^{1/2}$,
where $S_c\simeq \mu/2$,  and $\kappa$ now  
serves as UV cutoff length of the field theory.
To leading order in $\kappa/L,\kappa/L_T$, 
the hydrodynamic action (\ref{finala}) of a vortex
is $S_{el} = \mu \ln[{\rm min}(L,L_T)/2\kappa] + S^{\prime}(L/L_T)$,
where $L/L_T$
measures the anisotropy of this
finite-size 2D Kosterlitz-Thouless problem.
In particular, we find $S'(L_T\approx L)\simeq 0.11 \mu$, 
while in the opposite limits, $S'\simeq 2\mu/\pi$.  
Below $S'$ is taken into account as renormalization of  $S_c$.

The next step is to analyze a QPS gas, where
textbook analysis \cite{chaikin} leads to the picture
of an interacting Coulomb gas
of charges $\nu_i=\pm 1$, fugacity $y=e^{-S_c}$, and total charge zero.  
For $\mu>\mu^\ast\approx 2$, QPS's are confined
into neutral pairs, quasi-long range superconductivity is present,
but QPS's cause a finite resistance below $T_c$ \cite{zaikin1}. 
For $\mu=\mu^\ast$, QPS proliferation leads to a 
Kosterlitz-Thouless transition to the normal metallic state.
(Of course, here ``normal'' does not imply Fermi-liquid behavior.)
The transition temperature $T_c$ is therefore not $T_c^0$ but follows
from the condition $\mu(T_c)=\mu^\ast$. 
Equation (\ref{mu1}) then yields
\begin{equation}\label{tc1}
T_c/T_c^0 = \left[1- \mu^\ast/ \alpha_0 N\right]^{g_c/(g_c-1)}.
\end{equation}
This $T_c$ depression is normally rather weak, 
e.g.~for $N=31$, we obtain $T_c/T_c^0=0.97$, but for small $N$, 
the effect can be large. Furthermore, other mechanisms not included in
our model could act to effectively reduce $\alpha_0$ and
hence $T_c$, e.g.~disorder and 
heating effects \cite{zaikin1}, or the electromagnetic environment
\cite{blatter}.

The temperature dependence of the linear resistance $R(T)=V/I$ for
$T<T_c$ can be obtained by computing the voltage drop $V$ for applied
current $I$.  Expanding the vortex partition function
up to order $y^2$ and extracting the imaginary part of the 
free energy $F(I)$ using the Langer approach \cite{langer}, 
$\Gamma(\pm I)=-2{\rm Im} F(\pm I)$  
can be interpreted as the rate for a phase slip by $\pm 2\pi$ \cite{zaikin1}.
The average change in phase is then $\langle \dot\phi \rangle=
2\pi[\Gamma(I)-\Gamma(-I)]$, which from the Josephson relation implies
a voltage drop $V=\pi[\Gamma(I)-\Gamma(-I)]/e$.  Using $\epsilon=\pi \hbar
I/e$, the rate $\Gamma(\epsilon)$ follows 
for $L,L_T \gg \kappa$ but arbitrary $L/L_T$ in the form 
\[
\Gamma(\epsilon)=\frac{c_s^2 L y^2}{\kappa^4}
\int_{-L/2}^{L/2} dx \int_{-\infty}^\infty dt e^{i\epsilon t
- \mu [G(t+x/c_s)+G(t-x/c_s)]} ,
\]
where
$G(t) = \ln \left[(L_T/\kappa) \sinh(\pi T|t|)\right]
+ i (\pi/2) {\rm sgn}(t).$
To evaluate the rate $\Gamma(\epsilon)$ for arbitrary $L/L_T$,
we replace the boundaries for the $x$-integral by a soft
exponential cutoff, switch to integration variables
$t'=t-x/c_s$ and $t^{\prime \prime}=t+x/c_s$, 
and use the auxiliary relation
\[
e^{-c_s | t^{\prime\prime}- t'|/L}= \frac{c_s}{\pi L}
\int_{-\infty}^\infty ds \frac{e^{-is(t^{\prime\prime}-t')}}
{s^2+(c_s/L)^2}.
\]
It is now straightforward to carry out the $t',t^{\prime\prime}$
 time integrations, 
and some algebra yields the 
linear resistance 
\begin{eqnarray}\label{resis}
\frac{R}{R_q} &=& \left(  \frac{\pi y \Gamma(\mu/2)}{ \Gamma(\mu/2+1/2)}
\right)^2 \frac{\pi L}{\kappa} \left(\frac{L_T}{\kappa}\right)^{3-2\mu} 
\\ \nonumber
 &\times &  \int_{0}^\infty du \frac{2/\pi}{1+u^2}
\left|\frac{\Gamma(\mu/2+iu L_T/2L)} {\Gamma(\mu/2)} \right|^4 
\end{eqnarray}
in units of the resistance quantum $R_q=\pi \hbar/2e^2$.
Equation (\ref{resis}) leads to good agreement with experimental data
\cite{kociak,kasnew}; a detailed comparison will be given
in Ref.~\cite{future}.
For $L/L_T\gg 1$, the $u$-integral approaches unity, 
and hence $R\propto T^{2\mu-3}$,
while for $L/L_T\ll 1$, dimensional scaling arguments
give $R\propto T^{2\mu-2}$.  
The exponents are determined by the temperature-dependent
stiffness (\ref{mu1}).  While both
power-law behaviors have been
reported in Ref.~\cite{zaikin1}, Eq.~(\ref{resis}) describes the full
crossover for arbitrary $L/L_T$.  In Refs.~\cite{kociak,kasnew},
typical lengths were $L\approx 1 \mu$m, which indeed puts one into the
crossover regime $L_T\approx L$. 

To conclude, we have studied superconductivity
in carbon nanotube ropes, starting from a model of ballistic
SWNTs with attractive intra-tube interactions and 
inter-tube Josephson coupling.
We have constructed the Ginzburg-Landau theory including quantum
fluctuations. This allows for detailed predictions about the 
critical temperature $T_c$ and the QPS-induced resistance below $T_c$.
  If repulsive Coulomb interactions
can be screened off efficiently, our theory suggests that
superconductivity may survive down to only a few transverse channels
in clean nanotube ropes.  
--- We acknowledge useful discussions with A. Altland, H. Bouchiat,
F. Essler, and A. Tsvelik.
This work has been supported by the EU network DIENOW and
by the SFB-TR 12 of the DFG.


\begin{thebibliography}{24}
\expandafter\ifx\csname natexlab\endcsname\relax\def\natexlab#1{#1}\fi
\expandafter\ifx\csname bibnamefont\endcsname\relax
  \def\bibnamefont#1{#1}\fi
\expandafter\ifx\csname bibfnamefont\endcsname\relax
  \def\bibfnamefont#1{#1}\fi
\expandafter\ifx\csname citenamefont\endcsname\relax
  \def\citenamefont#1{#1}\fi
\expandafter\ifx\csname url\endcsname\relax
  \def\url#1{\texttt{#1}}\fi
\expandafter\ifx\csname urlprefix\endcsname\relax\def\urlprefix{URL }\fi
\providecommand{\bibinfo}[2]{#2}
\providecommand{\eprint}[2][]{\url{#2}}

\bibitem[{\citenamefont{Kociak et~al.}(2001)\citenamefont{Kociak, Kasumov,
  Gueron, Reulet, Khodos, Gorbatov, Volkov, Vaccarini, and Bouchiat}}]{kociak}
\bibinfo{author}{\bibfnamefont{M.}~\bibnamefont{Kociak}},
  \bibinfo{author}{\bibfnamefont{A.~Y.} \bibnamefont{Kasumov}},
  \bibinfo{author}{\bibfnamefont{S.}~\bibnamefont{Gueron}},
  \bibinfo{author}{\bibfnamefont{B.}~\bibnamefont{Reulet}},
  \bibinfo{author}{\bibfnamefont{I.~I.} \bibnamefont{Khodos}},
  \bibinfo{author}{\bibfnamefont{Y.~B.} \bibnamefont{Gorbatov}},
  \bibinfo{author}{\bibfnamefont{V.~T.} \bibnamefont{Volkov}},
  \bibinfo{author}{\bibfnamefont{L.}~\bibnamefont{Vaccarini}},
  \bibnamefont{and} \bibinfo{author}{\bibfnamefont{H.}~\bibnamefont{Bouchiat}},
  \bibinfo{journal}{Phys. Rev. Lett.} \textbf{\bibinfo{volume}{86}},
  \bibinfo{pages}{2416} (\bibinfo{year}{2001}).

\bibitem[{\citenamefont{Tang et~al.}(2001)\citenamefont{Tang, Zhang, Wang,
  Zhang, Wen, Li, Wang, Chan, and Sheng}}]{tang}
\bibinfo{author}{\bibfnamefont{Z.~K.} \bibnamefont{Tang}},
  \bibinfo{author}{\bibfnamefont{L.}~\bibnamefont{Zhang}},
  \bibinfo{author}{\bibfnamefont{N.}~\bibnamefont{Wang}},
  \bibinfo{author}{\bibfnamefont{X.~X.} \bibnamefont{Zhang}},
  \bibinfo{author}{\bibfnamefont{G.~H.} \bibnamefont{Wen}},
  \bibinfo{author}{\bibfnamefont{G.~D.} \bibnamefont{Li}},
  \bibinfo{author}{\bibfnamefont{J.~N.} \bibnamefont{Wang}},
  \bibinfo{author}{\bibfnamefont{C.~T.} \bibnamefont{Chan}}, \bibnamefont{and}
  \bibinfo{author}{\bibfnamefont{P.}~\bibnamefont{Sheng}},
  \bibinfo{journal}{Science} \textbf{\bibinfo{volume}{292}},
  \bibinfo{pages}{2462} (\bibinfo{year}{2001}).

\bibitem[{\citenamefont{Kasumov et~al.}(2003)\citenamefont{Kasumov, Kociak,
  Ferrier, Deblock, Gueron, Reulet, Khodos, Stephan, and Bouchiat}}]{kasnew}
\bibinfo{author}{\bibfnamefont{A.}~\bibnamefont{Kasumov}},
  \bibinfo{author}{\bibfnamefont{M.}~\bibnamefont{Kociak}},
  \bibinfo{author}{\bibfnamefont{M.}~\bibnamefont{Ferrier}},
  \bibinfo{author}{\bibfnamefont{R.}~\bibnamefont{Deblock}},
  \bibinfo{author}{\bibfnamefont{S.}~\bibnamefont{Gueron}},
  \bibinfo{author}{\bibfnamefont{B.}~\bibnamefont{Reulet}},
  \bibinfo{author}{\bibfnamefont{I.}~\bibnamefont{Khodos}},
  \bibinfo{author}{\bibfnamefont{O.}~\bibnamefont{Stephan}}, \bibnamefont{and}
  \bibinfo{author}{\bibfnamefont{H.}~\bibnamefont{Bouchiat}}
  (\bibinfo{year}{2003}), \eprint{cond-mat/0307260}.

\bibitem[{\citenamefont{Kasumov et~al.}(1999)\citenamefont{Kasumov, Deblock,
  Kociak, Reulet, Bouchiat, Khodos, Gorbatov, Volkov, Journet, and
  Burghard}}]{kasumov}
\bibinfo{author}{\bibfnamefont{A.~Y.} \bibnamefont{Kasumov}},
  \bibinfo{author}{\bibfnamefont{R.}~\bibnamefont{Deblock}},
  \bibinfo{author}{\bibfnamefont{M.}~\bibnamefont{Kociak}},
  \bibinfo{author}{\bibfnamefont{B.}~\bibnamefont{Reulet}},
  \bibinfo{author}{\bibfnamefont{H.}~\bibnamefont{Bouchiat}},
  \bibinfo{author}{\bibfnamefont{I.~I.} \bibnamefont{Khodos}},
  \bibinfo{author}{\bibfnamefont{Y.~B.} \bibnamefont{Gorbatov}},
  \bibinfo{author}{\bibfnamefont{V.~T.} \bibnamefont{Volkov}},
  \bibinfo{author}{\bibfnamefont{C.}~\bibnamefont{Journet}}, \bibnamefont{and}
  \bibinfo{author}{\bibfnamefont{M.}~\bibnamefont{Burghard}},
  \bibinfo{journal}{Science} \textbf{\bibinfo{volume}{284}},
  \bibinfo{pages}{1508} (\bibinfo{year}{1999}).

\bibitem[{\citenamefont{Morpurgo et~al.}(1999)\citenamefont{Morpurgo, Kong,
  Marcus, and Dai}}]{morpurgo}
\bibinfo{author}{\bibfnamefont{A.~F.} \bibnamefont{Morpurgo}},
  \bibinfo{author}{\bibfnamefont{J.}~\bibnamefont{Kong}},
  \bibinfo{author}{\bibfnamefont{C.~M.} \bibnamefont{Marcus}},
  \bibnamefont{and} \bibinfo{author}{\bibfnamefont{H.}~\bibnamefont{Dai}},
  \bibinfo{journal}{Science} \textbf{\bibinfo{volume}{286}},
  \bibinfo{pages}{263} (\bibinfo{year}{1999}).

\bibitem[{\citenamefont{Gonz{\'a}lez}(2002)}]{gonzalez1}
\bibinfo{author}{\bibfnamefont{J.}~\bibnamefont{Gonz{\'a}lez}},
  \bibinfo{journal}{Phys. Rev. Lett.} \textbf{\bibinfo{volume}{88}},
  \bibinfo{pages}{076403} (\bibinfo{year}{2002}).

\bibitem[{\citenamefont{Gonz{\'a}lez}(2003)}]{gonzalez2}
\bibinfo{author}{\bibfnamefont{J.}~\bibnamefont{Gonz{\'a}lez}},
  \bibinfo{journal}{Phys. Rev. B} \textbf{\bibinfo{volume}{67}},
  \bibinfo{pages}{014528} (\bibinfo{year}{2003}).

\bibitem[{\citenamefont{De~Martino and Egger}(2003{\natexlab{a}})}]{ademarti}
\bibinfo{author}{\bibfnamefont{A.}~\bibnamefont{De~Martino}} \bibnamefont{and}
  \bibinfo{author}{\bibfnamefont{R.}~\bibnamefont{Egger}},
  \bibinfo{journal}{Phys. Rev. B} \textbf{\bibinfo{volume}{67}},
  \bibinfo{pages}{235418} (\bibinfo{year}{2003}{\natexlab{a}}).

\bibitem[{\citenamefont{Schulz and Bourbonnais}(1983)}]{schulz}
\bibinfo{author}{\bibfnamefont{H.~J.} \bibnamefont{Schulz}} \bibnamefont{and}
  \bibinfo{author}{\bibfnamefont{C.}~\bibnamefont{Bourbonnais}},
  \bibinfo{journal}{Phys. Rev. B} \textbf{\bibinfo{volume}{27}},
  \bibinfo{pages}{5856} (\bibinfo{year}{1983}).

\bibitem[{\citenamefont{Lau et~al.}(2001)\citenamefont{Lau, Markovic, Bockrath,
  Bezryadin, and Tinkham}}]{lau}
\bibinfo{author}{\bibfnamefont{C.~N.} \bibnamefont{Lau}},
  \bibinfo{author}{\bibfnamefont{N.}~\bibnamefont{Markovic}},
  \bibinfo{author}{\bibfnamefont{M.}~\bibnamefont{Bockrath}},
  \bibinfo{author}{\bibfnamefont{A.}~\bibnamefont{Bezryadin}},
  \bibnamefont{and} \bibinfo{author}{\bibfnamefont{M.}~\bibnamefont{Tinkham}},
  \bibinfo{journal}{Phys. Rev. Lett.} \textbf{\bibinfo{volume}{87}},
  \bibinfo{pages}{217003} (\bibinfo{year}{2001}).

\bibitem[{\citenamefont{Zaikin et~al.}(1997)\citenamefont{Zaikin, Golubev, van
  Otterlo, and Zimanyi}}]{zaikin1}
\bibinfo{author}{\bibfnamefont{A.~D.} \bibnamefont{Zaikin}},
  \bibinfo{author}{\bibfnamefont{D.~S.} \bibnamefont{Golubev}},
  \bibinfo{author}{\bibfnamefont{A.}~\bibnamefont{van Otterlo}},
  \bibnamefont{and} \bibinfo{author}{\bibfnamefont{G.~T.}
  \bibnamefont{Zimanyi}}, \bibinfo{journal}{Phys. Rev. Lett.}
  \textbf{\bibinfo{volume}{78}}, \bibinfo{pages}{1552} (\bibinfo{year}{1997}).

\bibitem[{\citenamefont{Golubev and Zaikin}(2001)}]{zaikin2}
\bibinfo{author}{\bibfnamefont{D.~S.} \bibnamefont{Golubev}} \bibnamefont{and}
  \bibinfo{author}{\bibfnamefont{A.~D.} \bibnamefont{Zaikin}},
  \bibinfo{journal}{Phys. Rev. B} \textbf{\bibinfo{volume}{64}},
  \bibinfo{pages}{014504} (\bibinfo{year}{2001}).

\bibitem[{\citenamefont{Tinkham}(1996)}]{tinkham}
\bibinfo{author}{\bibfnamefont{M.}~\bibnamefont{Tinkham}},
  \emph{\bibinfo{title}{Introduction to Superconductivity, 2nd Edition}}
  (\bibinfo{publisher}{McGraw-Hill, Inc.}, \bibinfo{year}{1996}).

\bibitem[{\citenamefont{B{\"u}chler et~al.}(2003)\citenamefont{B{\"u}chler,
  Geshkenbein, and Blatter}}]{blatter}
\bibinfo{author}{\bibfnamefont{H.~P.} \bibnamefont{B{\"u}chler}},
  \bibinfo{author}{\bibfnamefont{V.~B.} \bibnamefont{Geshkenbein}},
  \bibnamefont{and} \bibinfo{author}{\bibfnamefont{G.}~\bibnamefont{Blatter}}
  (\bibinfo{year}{2003}), \eprint{cond-mat/0306617}.

\bibitem[{\citenamefont{Egger and Gogolin}(1997)}]{egger97}
\bibinfo{author}{\bibfnamefont{R.}~\bibnamefont{Egger}} \bibnamefont{and}
  \bibinfo{author}{\bibfnamefont{A.~O.} \bibnamefont{Gogolin}},
  \bibinfo{journal}{Phys. Rev. Lett.} \textbf{\bibinfo{volume}{79}},
  \bibinfo{pages}{5082} (\bibinfo{year}{1997}).

\bibitem[{\citenamefont{Kane et~al.}(1997)\citenamefont{Kane, Balents, and
  Fisher}}]{kane97}
\bibinfo{author}{\bibfnamefont{C.} \bibnamefont{Kane}},
  \bibinfo{author}{\bibfnamefont{L.}~\bibnamefont{Balents}}, \bibnamefont{and}
  \bibinfo{author}{\bibfnamefont{M.~P.~A.} \bibnamefont{Fisher}},
  \bibinfo{journal}{Phys. Rev. Lett.} \textbf{\bibinfo{volume}{79}},
  \bibinfo{pages}{5086} (\bibinfo{year}{1997}).

\bibitem[{\citenamefont{Maarouf et~al.}(2000)\citenamefont{Maarouf, Kane, and
  Mele}}]{kane}
\bibinfo{author}{\bibfnamefont{A.~A.} \bibnamefont{Maarouf}},
  \bibinfo{author}{\bibfnamefont{C.~L.} \bibnamefont{Kane}}, \bibnamefont{and}
  \bibinfo{author}{\bibfnamefont{E.~J.} \bibnamefont{Mele}},
  \bibinfo{journal}{Phys. Rev. B} \textbf{\bibinfo{volume}{61}},
  \bibinfo{pages}{11156} (\bibinfo{year}{2000}).

\bibitem[{\citenamefont{Carr and Tsvelik}(2002)}]{carr}
\bibinfo{author}{\bibfnamefont{S.~T.} \bibnamefont{Carr}} \bibnamefont{and}
  \bibinfo{author}{\bibfnamefont{A.~M.} \bibnamefont{Tsvelik}},
  \bibinfo{journal}{Phys. Rev. B} \textbf{\bibinfo{volume}{65}},
  \bibinfo{pages}{195121} (\bibinfo{year}{2002}).

\bibitem[{\citenamefont{Lukyanov and Zamolodchikov}(1997)}]{zamo}
\bibinfo{author}{\bibfnamefont{S.}~\bibnamefont{Lukyanov}} \bibnamefont{and}
  \bibinfo{author}{\bibfnamefont{A.~B.} \bibnamefont{Zamolodchikov}},
  \bibinfo{journal}{Nucl. Phys. B} \textbf{\bibinfo{volume}{493}},
  \bibinfo{pages}{571} (\bibinfo{year}{1997}).

\bibitem[{\citenamefont{Nagaosa}(1999)}]{nagaosa1}
\bibinfo{author}{\bibfnamefont{N.}~\bibnamefont{Nagaosa}},
  \emph{\bibinfo{title}{Quantum Field Theory in Condensed Matter Physics}}
  (\bibinfo{publisher}{Springer Verlag}, \bibinfo{year}{1999}).

\bibitem[{\citenamefont{Mooij and Sch{\"o}n}(1985)}]{mooji}
\bibinfo{author}{\bibfnamefont{J.~E.} \bibnamefont{Mooij}} \bibnamefont{and}
  \bibinfo{author}{\bibfnamefont{G.}~\bibnamefont{Sch{\"o}n}},
  \bibinfo{journal}{Phys. Rev. Lett.} \textbf{\bibinfo{volume}{55}},
  \bibinfo{pages}{114} (\bibinfo{year}{1985}).

\bibitem[{\citenamefont{De~Martino and Egger}(2003{\natexlab{b}})}]{future}
\bibinfo{author}{\bibfnamefont{A.}~\bibnamefont{De~Martino}} \bibnamefont{and}
  \bibinfo{author}{\bibfnamefont{R.}~\bibnamefont{Egger}}
  (\bibinfo{year}{2003}{\natexlab{b}}), \eprint{in preparation}.

\bibitem[{\citenamefont{Chaikin and Lubensky}(2000)}]{chaikin}
\bibinfo{author}{\bibfnamefont{P.~M.} \bibnamefont{Chaikin}} \bibnamefont{and}
  \bibinfo{author}{\bibfnamefont{T.}~\bibnamefont{Lubensky}},
  \emph{\bibinfo{title}{Principles of Condensed Matter Physics}}
  (\bibinfo{publisher}{Cambridge University Press}, \bibinfo{year}{2000}).

\bibitem[{\citenamefont{Langer}(1967)}]{langer}
\bibinfo{author}{\bibfnamefont{J.~S.} \bibnamefont{Langer}},
  \bibinfo{journal}{Ann. Phys. (N.Y.)} \textbf{\bibinfo{volume}{41}},
  \bibinfo{pages}{108} (\bibinfo{year}{1967}).

\end{thebibliography}
\end{document}